\documentclass[12pt,onecolumn,preprintnumbers,amsmath,amssymb]{revtex4}
\usepackage{graphicx}% Include figure files
\usepackage{dcolumn}% Align table columns on decimal point
\usepackage{bm}% bold math
\usepackage{setspace}
\usepackage{color}
\usepackage{appendix}

\begin{document}

%\preprint{APS/123-QED}

\title{Material properties of \emph{Caenorhabditis elegans} swimming at low Reynolds number}% Force line breaks with \\

\author{J. Sznitman$^{1}\footnote{Present address: Department of Mechanical $\&$ Aerospace Engineering, Princeton University, Princeton NJ 08544, USA}$}
\author{Prashant K. Purohit$^1$}%
\author{P. Krajacic$^2$}
\author{T. Lamitina$^2$}
\author{P.E. Arratia$^1$}\email{parratia@seas.upenn.edu}
\affiliation{%
$^1$Department of Mechanical Engineering \& Applied Mechanics, University of Pennsylvania, Philadelphia, PA 19104, USA}
\affiliation{$^2$Department of Physiology,University of Pennsylvania, Philadelphia, PA 19104, USA}

%\affiliation{

%\singlespacing
\begin{abstract}
\textbf{Abstract:} Undulatory locomotion, as seen in the nematode \emph{Caenorhabditis elegans}, is a common swimming gait of organisms in the low Reynolds number regime, where viscous forces are dominant. While the nematode's motility is expected to be a strong function of its material properties, measurements remain scarce. Here, the swimming behavior of \emph{C.} \emph{elegans} are investigated in experiments and in a simple model. Experiments reveal that nematodes swim in a periodic fashion and generate traveling waves which decay from head to tail. The model is able to capture the experiments' main features and is used to estimate the nematode's Young's modulus $E$ and tissue viscosity $\eta$. For wild-type \emph{C. elegans}, we find $E\approx 3.77$~kPa and $\eta \approx-860$~Pa$\cdot$s; values of $\eta$ for live \emph{C. elegans} are negative because the tissue is generating rather than dissipating energy. Results show that material properties are sensitive to changes in muscle functional properties, and are useful quantitative tools with which to more accurately describe new and existing muscle mutants.

\end{abstract}

\pacs{Valid PACS appear here}% PACS, the Physics and Astronomy
\maketitle

\section{\label{sec:level1}Introduction}

Motility analysis of model organisms, such as the nematode \emph{Caenorhabditis elegans} (\emph{C. elegans}), is of great scientific and practical interest. It can provide, for example, a powerful tool for the analysis of genetic diseases in humans such as muscular dystrophy (MD)~\cite{bargmann1998,mendel1995,nelson1998} since \emph{C. elegans} have muscle cells that are highly similar in both anatomy and molecular makeup to vertebrate skeletal muscles \cite{white1976,white1986}. Due to the nematode's small size ($L\approx1$~mm), the motility of \emph{C. elegans} swimming in a simple, Newtonian fluid is usually investigated in the low Reynolds numbers ($Re$) regime, where linear viscous forces dominate over nonlinear inertial forces \cite{childress1981,brennen1977}. At low $Re$, locomotion results from non-reciprocal deformations to break time-reversal symmetry \cite{taylor1951}; this is the so-called ``scallop theorem'' \cite{purcell1977}. Experimental observations have shown that motility of swimming nematodes including \emph{C. elegans} results from the propagation of bending waves along the nematode's body length \cite{gray1964,korta2007,pierce2008}. These waves consist of alternating phases of dorsal and ventral muscle contractions driven by the neuromuscular activity of muscle cells. While it is generally accepted that during locomotion the nematode's tissues obey a viscoelastic reaction \cite{karbow2006,guo2008,fung1993}, quantitative data on \emph{C. elegans'} material properties such as tissue viscosity and Young's modulus remain largely unexplored.

Motility behavior of \emph{C. elegans} is a strong function of its body material properties. Recent investigations have provided valuable data on \emph{C. elegans'} motility, such as velocity, bending frequency, and body wavelength~\cite{cronin2005,feng2004,karbow2006,korta2007,pierce2008,ramot2008}. However, only recently have the nematode's material properties been probed using piezoresistive cantilevers \cite{park2007}. Such invasive measurements provided Young's modulus values of the \emph{C. elegans'} cuticle on the order of 400 MPa; this value is closer to stiff rubber than to soft tissues.

In this paper, we investigate the motility of \emph{C. elegans} in both experiments and in a model in order to estimate the nematode's material properties. Experiments show that nematodes swim in a highly periodic fashion and generate traveling waves which decay from head to tail. A dynamic model is proposed based on force and moment (torque) balance. A simplified version of the model is able to capture the main features of the experiments such as the traveling waves and their decay. The model is used to estimate both the Young's modulus and tissue viscosity of \emph{C. elegans}. Such estimates are used to characterize motility phenotypes of healthy nematodes and mutants carrying muscular dystrophy (MD).

\section{\label{sec:level1}Experimental Methods}
Experiments are performed by imaging \emph{C. elegans} using standard microscopy and a high-speed camera at 125 frames per second. We focus our analysis on forward swimming in shallow channels to minimize three-dimensional motion. Channels are machined in acrylic and are $1.5$~mm wide and $500$~$\mu$m deep; they are sealed with a thin ($0.13$~mm) cover glass. Channels are filled with an aqueous solution of M9 buffer \cite{brenner1974}, which contains $5$ to $10$ nematodes. The buffer viscosity $\mu$ and density $\rho$ are 1.1 cP and 1.0 g/cm$^3$, respectively. Under such conditions, the Reynolds number, defined as $Re=\rho U L/\mu$, is less than unity, where $U$ and $L$ are the nematode's swimming speed and length, respectively.

In Fig.~\ref{fig_1}(a), we display nematode tracking data over multiple bending cycles for a healthy, wild-type nematode. Results show that the nematode swims with an average speed $<U>=0.45$~mm/s and with a beating pattern of period $T=0.46$~s. This periodic behavior is also qualitatively observed in the motion of the nematode tail (Fig.~\ref{fig_1}a); see also Video 1 (Supplementary Material). Under such conditions, $Re \approx 0.4$. Snapshots of the nematode skeletons over one beating cycle (Fig.~\ref{fig_1}b) reveal an envelope of well-confined body postures with a wavelength of approximately $1$~mm, which corresponds nearly to the nematode's body length. The displacement amplitudes at the head and tail are similar with $465~\mu$m and $400~\mu$m, respectively. However, the amplitudes of the curvature at the head and tail differ sharply with approximately $6.07~$mm$^{-1}$ and $2.21~$mm$^{-1}$, respectively. The tail/head curvature ratio of $0.36$ suggests that the bending motion is initiated at the head.

Extensive genetic analysis in \emph{C. elegans} has identified numerous mutations affecting nematode motility. One such mutant, \emph{dys-1}, encodes a homolog of the human dystrophin protein, which is mutated in Duchenne's and Becker's muscular dystrophy (MD).  Using qualitative observation, \emph{dys-1} mutants have an extremely subtle movement defect \cite{gieseler2000}, which includes slightly exaggerated head bending and time-dependent decay in movement. The quantitative imaging platform presented here is able to robustly differentiate between wild type and \emph{dys-1} mutants, as shown in Fig.~\ref{fig_1}(c) and Video 2 (Supplementary Material).

Results show that the \emph{dys-1} mutant swims with an average speed $<U>=0.17$~mm/s and $Re~\approx~0.15$; both values are significantly smaller than for the wild-type nematode. Although the \emph{dys-1} mutant suffers from severe motility defects~\cite{gieseler2000}, it still moves in a highly periodic fashion with $T=0.63$~s.  Snapshots of nematode skeletons over one beating cycle (Fig.~\ref{fig_1}d) also reveal an envelope of well-confined body postures with a wavelength corresponding to the nematode's body length. The \emph{dys-1} mutant exhibits a tail/head curvature ratio of $0.23$, which is similar to the value found for wild type nematodes. However, the corresponding displacement amplitudes of the mutant are much smaller than those observed for the wild-type (Fig.~\ref{fig_1}b). The displacement amplitudes at the head and tail are $330~\mu$m and $155~\mu$m, respectively. This observation suggests that the \emph{dys-1} mutant body, and in particular the tail, are becoming inactive as the bending motion is not able to deliver as much body displacement.

To further characterize the motility of \emph{C. elegans}, we measure the curvature $\kappa(s,t)=d \phi / ds$ along the nematode's body (Fig.~\ref{fig_2}a). Here, $\phi$ is the angle made by the tangent to the $x$-axis at each point along the centerline and $s$ is the arc-length coordinate spanning the nematode's head ($s=0$) to its tail ($s=L$). The spatio-temporal evolution of $\kappa$ for a swimming nematode is shown in Fig.~\ref{fig_2}(a). Approximately $6$ bending cycles are illustrated and curvature values are color-coded; red and blue represent positive and negative values of $\kappa$, respectively. The $y$-axis in Fig.~\ref{fig_2}(a) corresponds to the non-dimensional body position $s/L$. The contour plot shows the existence of highly periodic, well-defined diagonally oriented lines. These diagonal lines are characteristic of bending waves, which propagate in time along the body length. Note that as the wave travels along the nematode body, \emph{the magnitude of $\kappa$ decays from head to tail}. Such behavior contrasts sharply with that observed for undulatory swimmers of the inertial regime (e.g. eel, lamprey), where amplitudes of body displacement grow instead from head to tail \cite{tytell2004,kern2006}.

The body bending frequency ($f$) is obtained from the one-dimensional Fast Fourier Transform (FFT) of the curvature field $\kappa$ at multiple body positions $s/L$ (Fig.~\ref{fig_2}b). Here, the body bending frequency is defined as $f= \omega/ 2 \pi$. The angular frequency $\omega$ is calculated by first extracting multiple lines from the curvature field at distinct body positions $s/L$, and then computing the one-dimensional FFT. The wave speed $c$ is extracted from the slope of the curvature $\kappa$ propagating along the nematode's body; the wavelength $\lambda$ is computed from the expression $\lambda = c/f$. A single frequency peak $f=2.17 \pm 0.18$~Hz ($n = 25$) is found in the Fourier spectrum, where $n$ is the number of nematodes. This single peak is irrespective of body position and corresponds to a wave speed $c = 2.14 \pm 0.16$~mm/s.

\section{\label{sec:level1} Mathematical Methods}
The swimming motion of \emph{C. elegans} is modeled as a slender body in the limit of low Reynolds numbers (\emph{Re})~\cite{qian2008,yu2006,grayhancock1955}. This model is later used to estimate the material properties of {\it C. elegans}. We assume that the nematode is inextensible~\cite{grayhancock1955}; the uncertainty in the measured body lengths is less than $3\%$. The nematode's motion is restricted to the $xy$-plane and is described in terms of its center-line $y(s,t)$, where $s$ is the arc-length along the filament~\cite{antman1995}. The swimming {\it C. elegans} experiences no net total force or torque (moments) such that, in the limit of low $Re$, the dynamic equations of motion are
\begin{eqnarray}
 \frac{\partial \vec{F}}{\partial s} & =& C_t \vec{u}_t
 + C_n \vec{u}_n, \label{force} \\
 \frac{\partial M}{\partial s}& =&-[F_{y}\cos(\phi)-F_{x}\sin(\phi)] \label{moment}.
\end{eqnarray}

In Eq.~(\ref{force}), $\vec{F}(s,t)$ is the internal force in the nematode, $C_{i}$ is the drag coefficient experienced by the nematode, $\vec{u_{i}}$ is the nematode velocity, and the subscripts $t$ and $n$ correspond to the tangent and normal directions, respectively. The drag coefficients $C_{t}$ and $C_{n}$ are obtained from slender body theory~\cite{grayhancock1955}. Due to the finite confinement of nematodes between parallel walls, corrections for wall effects on the resistive coefficients are estimated for slender cylinders \cite{brennen1977,katz1975}.

In Eq.~(\ref{moment}), $M=M_{p}+M_{a}$, where $M_{p}$ is a passive moment and $M_{a}$ is an active moment generated by the muscles of the nematode; the active and passive moments are parts of a total internal moment~\cite{karbow2006,guo2008}. The passive moment is given by the Voigt model \cite{fung1993} such that $M_{p} = EI\kappa + \eta_{p} I(\partial\kappa/\partial t)$, where $I$ is the second moment of inertia of the nematode cross section. The Voigt model is one of the simplest models for muscle and is extensively used in the literature \cite{linden1974}. Qualitatively, the elastic part of the Voigt model is represented by a spring of stiffness $E$ while the dissipative part of the Voigt model is represented by a dashpot filled with a fluid of viscosity $\eta_p$ (Fig.~\ref{fig_3}). Here, we assume two homogeneous {\it effective} material properties, namely (i) a constant Young's modulus $E$ and (ii) a constant tissue viscosity $\eta_{p}$.

The active moment generated by the muscle is given by $M_{a} = -(EI\kappa_{a} + \eta_{a}I \partial\kappa/\partial t)$, where $\kappa_{a}$ is a space and time dependent preferred curvature produced by the muscles of the nematode and $\eta_{a}$ is a positive constant~\cite{thomas1998}. A simple form for $\kappa_{a}$ can be obtained by assuming that $\kappa_{a}$ is a sinusoidal function of time with an amplitude that decreases
from the nematode's head to its tail (see Appendix). Note that if $\eta = \eta_{p} - \eta_{a} > 0$, there is net dissipation of energy in the tissue; conversely, if $\eta = \eta_{p} - \eta_{a} < 0$, there is net generation of energy in the tissue. Experiments have shown that the force generated by active muscle decreases with increasing velocity of shortening \cite{linden1974}, so that the force-velocity curve for active muscle has a negative slope (Fig.~\ref{fig_3}). Such \emph{negative} viscosity has been derived by a mathematical analysis of the kinetics of the mechano-chemical reactions in the cross-bridge cycle of active muscles \cite{thomas1998}. For live nematodes, we expect $\eta = \eta_{p} - \eta_{a} < 0$ because the net energy produced in the (muscle) tissue is needed to overcome the drag from the surrounding fluid.

Equations~(\ref{force}) and (\ref{moment}) are simplified by noting that the nematode moves primarily in the $x$-direction (see Video 1, S.M.) and that the deflections of its centerline from the $x$-axis are small. In such case, $s \approx x$ and $\cos(\phi)\approx 1$. This results in a linearized set of equations given by
\begin{eqnarray}
\frac{\partial F_{y}}{\partial x}-C_{n}\frac{\partial y}{\partial t}=0,
\label{eq:yforbal} \\
\frac{\partial M}{\partial x}+F_{y}=0. \label{eq:mombal}
\end{eqnarray}
Differentiating Eq.~(\ref{eq:mombal}) with respect to $x$ and combining with Eq.~(\ref{eq:yforbal}), we obtain
\begin{equation}
\frac{\partial^{2} M}{\partial x^{2}}+C_{n}\frac{\partial y}{\partial t}=0. \label{eq:final}
\end{equation}
Substituting for $M(x,t)$ in terms of $\kappa(x,t)$ and its
time derivative yields a bi-harmonic equation for the displacement $y(x,t)$ of the type
\begin{equation}
\frac{\partial^{4} y}{\partial x^{4}}+ \xi\frac{\partial y}{\partial t}=0, \label{eq:biharmonic}
\end{equation}
which can be solved analytically for appropriate boundary conditions, where $\xi$ is a constant that depends on the nematode's material properties and the fluid drag coefficient (see Appendix).

The boundary conditions are such that both the force and moment at the nematode's head and tail are equal to zero. That is, $F_{y}(0,t)=F_{y}(L,t)=0$ and $M(0,t)=M(L,t)=0$. Note that the zero moment boundary conditions at the head and tail imply that $EI\kappa(0,t) + \eta I(\partial\kappa(0,t)/\partial t) = EI\kappa_{a}(0,t)$ and $EI\kappa(L,t) + \eta I(\partial\kappa(L,t)/\partial t) = EI\kappa_{a}(L,t)$.

Experiments show that the curvature $\kappa$ has non-zero amplitudes (Fig.~\ref{fig_2}a) both at the head ($x=0$) and the tail ($x=L$). In order to capture this observation, we assume that $\kappa_{a}(x,t)$ is a sinusoidal wave with decreasing amplitude of the form $\kappa_{a}(x,t) = Q_{0}\cos\omega t + Q_{1}x\cos(\omega t - B)$, where $Q_{0}$, $Q_{1}$ and $B$ are inferred from the experiments. Note that if the curvature amplitude at the head is larger than that at the tail, then the nematode swims forward. Conversely, if the curvature amplitude is smaller at the head than at the tail, the nematode swims backward; if the amplitudes are equal at the head and tail then it remains stationary. We note, however, that other forms of the preferred curvature $\kappa_{a}$ are possible and could replicate the behavior seen in experiments.

\section{\label{sec:level1} Results \& Discussion}
Equation~(\ref{eq:biharmonic}) is solved for the displacement $y(x,t)$ in
order to obtain the curvature $\kappa(x,t) = \partial^{2}y/\partial x^{2}$.
The solution for $y(x,t)$ is a superposition of four traveling waves of the
general form $A_{i}\exp(-\beta x\cos P_{i})\cos(\beta x\sin P_{i}-\omega t -
\phi_{i})$ where $\beta = (C_n \omega /K_b)^{1/4}$ and $P_{i}$
is a function of the phase angle $\psi$. The amplitude $A_{i}$ and phase
$\phi_{i}$ are constants to be determined by enforcing the boundary conditions
discussed above (see Appendix). The solution reveals both the traveling
bending wave and the characteristic decay in $\kappa$, as seen in experiments.
Note that our formulation does not assume a wave-functional
form for $\kappa(x,t)$. Rather the wave is obtained as part of the solution.

Next, the curvature amplitude $|\kappa(x)|$ predicted by the model is fitted to those obtained from experiment to estimate the bending modulus $K_{b} = I\sqrt{E^{2} + \omega^{2}\eta^{2}}$ and the phase angle $\psi = \tan^{-1}(\eta\omega)/E$. The nematode is assumed to be a hollow, cylindrical shell~\cite{park2007,zelenskaya2005} such that $I = \pi((r_{m}+t/2)^{4} - (r_{m}-t/2)^{4})/4$, where the mean nematode radius $r_{m} \approx 35~\mu$m and the cuticle thickness $t\approx 0.5~\mu$m~\cite{cox1981}. For the population of wild-type \emph{C. elegans} tested here ($n=25$), the best fit values are $K_b = 4.19 \times 10^{-16} \pm 0.49 \times 10^{-16}$~Nm$^2$ and $\psi=-45.3^o \pm 3.0^o$. In Fig.~\ref{fig_4}, the experimental values of $|\kappa(x)|$ along the body of a wild-type \emph{C. elegans} are displayed together with theoretical  values of $|\kappa(x)|$, which are obtained by using the best fit value of the bending modulus for this nematode and by changing the phase angle from $\psi=0^o$ to $\psi=-90^o$. Figure~\ref{fig_4} shows that the model is able to capture the decay in $|\kappa|$ as a function of body length and the nematode's viscoelastic behavior.

The values of Young's modulus $E$ and tissue viscosity $\eta$ can now be estimated based on the values of $\psi$ and $K_{b}$ discussed above. Results show that, for the wild-type nematodes, $E=3.77 \pm 0.62$~kPa and $\eta = -860.2 \pm 99.4$~Pa$\cdot$s. The estimated value of $E$ lies in the range of values of tissue elasticity measured for isolated brain ($0.1-1~$kPa) and muscle cells ($8-17~$kPa) \cite{engler2006}. The values of $\eta$ for live \emph{C. elegans} are negative because the tissue is generating rather than dissipating energy~\cite{thomas1998,feit1985,kawai1980,tawada1990}. We note, however, that the absolute values of tissue viscosity $|\eta|$ are within the range ($10^2-10^4~$Pa$\cdot$s) measured for living cells~\cite{yamada2000,thoumine1997}. In order to determine whether the nematode's material properties can be extracted reliably from shape measurements alone, experiments in solutions of different viscosities were conducted~\cite{purohit2009}. The inferred Young's modulus and effective tissue viscosity remain constant for up to a $5$-fold increase in the surrounding fluid viscosity (or mechanical load).

The nematode's curvature $\kappa(x,t)$ is now determined from the estimated values of $E$ and $\eta$. Figure~\ref{fig_2}(c) shows a typical curvature $\kappa(x,t)$ contour plot obtained from the solution of the above equations using the estimated values of $E$ and $\eta$. While the influence of non-linearities is neglected for the scope of the present paper, the analytical results show that our linearized model, while not perfect, is nevertheless able to capture the main features observed in experiments (Fig.~\ref{fig_2}d).

Next, the method described above is used to quantify motility phenotypes of three distinct mutant MD strains (see Table 1 in Supplementary Material): one with a well-characterized muscle defect (\emph{dys-1;hlh-1}); one with a qualitatively subtle movement defect (\emph{dys-1}), and one mutant that has never been characterized with regards to motility phenotypes but is homologous to a human gene that causes a form of muscular dystrophy expressed in nematode muscle (\emph{fer-1}). Note that while both \emph{fer-1} and \emph{dys-1} genes are expressed in \emph{C. elegans} muscle, they exhibit little, if any, change in whole nematode motility under standard lab assays~\cite{bessou1998}.

Figure~\ref{fig_5} displays results of both kinematics (\ref{fig_5}a) and tissue material properties (\ref{fig_5}b) for all nematodes investigated here. Quantitative results are summarized in Table 1 (Supplementary Material). We find that all three mutants exhibit significant changes in both motility kinematics and tissue properties. For example, \emph{fer-1} mutants exhibit defects in motility kinematics which are not found with standard assays~\cite{bessou1998}. Specifically, both the maximum amount of body curvature attained in \emph{fer-1} mutants is increased by $\sim 5\%$, and the rate of curvature decay along the body is increased by $\sim 5\%$ (Fig.~\ref{fig_5}a). This data show that \emph{fer-1(hc24)} mutants exhibit small yet noteworthy defects in whole nematode motility and exhibit an \emph{uncoordinated} (\emph{unc}) phenotype. In comparison, kinematics data on \emph{dys-1;hlh-1} show that body curvature at the head of such mutant nematodes increases by $\sim 70\%$ compared to wild-type nematodes, while the rate of decay along the body is increased by approximately $\sim 40\%$. These results are useful to quantify the paralysis seen earlier in the tail motion of such MD mutants (Fig.~\ref{fig_1}c).

The Young's modulus ($E$) and the absolute values of tissue viscosity ($|\eta|$) of wild-type and mutant strains are shown in Fig.~\ref{fig_5}(b). Results show that mutants have lower values of $E$ when compared to wild-type nematodes. In other words, \emph{dys-1}, \emph{dys-1;hlh-1}, and \emph{fer-1} mutants \emph{C. elegans} are softer than their wild-type counterpart. The values of $|\eta|$ of \emph{fer-1} mutants are similar to wild-type nematodes, within experimental error. However, the values of $|\eta|$ for \emph{dys-1} mutants are lower than wild-type \emph{C. elegans}. Since muscle fibers are known to exhibit visible damage for \emph{dys-1;hlh-1} mutants~\cite{gieseler2000}, we hypothesize that the deterioration of muscle fibers may be responsible for the lower values of $E$ and $|\eta|$ found for \emph{dys-1} and \emph{dys-1;hlh-1} mutants.

\section{\label{sec:level1} Conclusion}
In summary, we characterize the swimming behavior of \emph{C. elegans} at low $Re$. Results show a distinct periodic swimming behavior with a traveling wave that decays from the nematode's head to tail. By coupling experiments with a linearized model based on force and torque balance, we are able to estimate, \emph{non-invasively}, the nematode's tissue material properties such as Young's modulus ($E$) and viscosity ($\eta$) as well as bending modulus ($K_{b}$). Results show that \emph{C. elegans} behaves effectively as a viscoelastic material with $E\approx3.77$~kPa, $|\eta|\approx860.2$~Pa$\cdot$s, and $K_b\approx4.19 \times 10^{-16}$~Nm$^2$. In particular, the estimated values of $E$ are much closer to biological tissues than previously reported values obtained using piezoresistive cantilevers~\cite{park2007}. We demonstrate that the methods presented here may be used, for example, to quantify motility phenotypes and tissue properties associated with muscular dystrophy mutations in \emph{C. elegans}. Overall, by combining kinematic data with a linearized model, we are able to provide a robust and highly quantitative phenotyping tool for analysis of \emph{C. elegans} motility, kinematics, and tissue mechanical properties.   Given the rapid non-invasive optical nature of this method, it may provide an ideal platform for genetic and small molecule screening applications aimed at correcting phenotypes of mutant nematodes. Our method also sheds new light on our understanding of muscle function, physiology, and animal locomotion in general.

\appendix*
\section{\label{sec:level1}}
In this appendix, we detail the model for the motion of the nematode
{\it C. elegans}. The
nematode is modeled as a slender filament at low Reynolds numbers~\cite{grayhancock1955}. In our experiments, the uncertainty in the measured
body lengths is less than $3\%$ and we assume inextensibility. The
nematode's motion is described in terms of its center-line $\vec{y}(s,t)$,
where $s$ is the arc-length along the filament and $t$ is
time \cite{antman1995}. We assume that the nematode moves in the $xy$-plane.
The swimming {\it C. elegans} experiences no net total force or
torque (moments) such that, in the limit of low $Re$, the
equations of motion are
\begin{eqnarray}
 \frac{\partial \vec{F}}{\partial s} & =& C_t \vec{u}_t
 + C_n \vec{u}_n, \label{force_app} \\
 \frac{\partial M}{\partial s}\vec{e}_{z} & =&
 -\hat{t}\times\vec{F} \label{moment_app},
\end{eqnarray}
where $\hat{t}(s,t) = \partial \vec{y}/\partial s$ is the tangent vector to
the center-line, $\vec{F}(s,t)$ is the internal force,
and $M(s,t)\vec{e}_{z} = \vec{M}(s,t)$ is the internal moment consisting of a
passive and active part \cite{guo2008, karbow2006}.
Tangential and normal velocities are respectively given by $\vec{u}_{t} = (\partial \vec{y}/\partial t\cdot \hat{t})\hat{t}$ and $\vec{u}_{n} =(\mathbf{I} - \hat{t}\otimes\hat{t})\partial \vec{y}/\partial t$.
The drag coefficients, $C_{t}$ and $C_{n}$, are obtained from slender body
theory \cite{grayhancock1955}. Due to the finite confinement of nematodes
between the parallel walls, corrections for wall effects on the resistive
coefficients are estimated for slender cylinders \cite{brennen1977}.
The local body position $\vec{y}$, the velocity at any body position
$\partial \vec{y}/ \partial t$, and the tangent vector $\hat{t}$ are
all experimentally measured.

The constitutive relation for the moment $M(s,t)$ in our inextensible filament
is assumed to be given by
\begin{equation}
 M = M_{p} + M_{a},
\end{equation}
where $M_{p}(s,t)$ is a passive moment and $M_{a}(s,t)$ is an active moment
generated by the muscles of the nematode. The passive moment is given by
a viscoelastic Voigt model \cite{fung1993}
\begin{equation}
M_{p} = EI\kappa + \eta_{p} I\frac{\partial\kappa}{\partial t},
\end{equation}
where $\kappa(s,t)$ is the curvature along the nematode. Here, we assume
two homogeneous {\it effective} material properties, namely (i) a constant
Young's modulus $E$ and (ii) a constant tissue viscosity $\eta_{p}$. The
active moment generated by the muscle is assumed to be given by:
\begin{equation}
 M_{a} = -(EI\kappa_{a}
+ \eta_{a}I\frac{\partial\kappa}{\partial t}),
\end{equation}
where $\kappa_{a} = Q_{0}\cos\omega t + Q_{1}s\cos(\omega t - B)$ is a preferred curvature and
$\eta_{a}$ is a positive constant~\cite{thomas1998}. $Q_{0}$, $Q_{1}$ and $B$
must be obtained by fitting to experiments.
$Q_{1}^{2}L^{2} + 2Q_{0}Q_{1}L\cos (B) < 0$ means that the curvature amplitudes at the
head are larger than those at the tail and we should expect traveling waves going from
head to tail causing the nematode to swim forward; similarly,
$Q_{1}^{2}L^{2} + 2Q_{0}Q_{1}L\cos (B) > 0$ should give traveling waves
going from tail to head so that the nematodes swim backward \cite{pierce2008}.
The total moment $M(s,t)$ can be written as
\begin{equation} \label{eq:constlaw}
 M = M_{p} + M_{a} = \textcolor{blue}{EI(\kappa - \kappa_{a})
+ (\eta_{p} - \eta_{a})I\frac{\partial\kappa}{\partial t}
 = EI(\kappa - \kappa_{a}) + \eta I \frac{\partial\kappa}{\partial t}}.
\end{equation}
Note that if $\eta = \eta_{p} - \eta_{a} > 0$ then there is net
dissipation of
energy in the tissue; if $\eta = \eta_{p} - \eta_{a} < 0$ then there is net
generation of energy in the tissue. For live nematodes that actively swim
in the fluid we expect $\eta = \eta_{p} - \eta_{a} < 0$ since the net
energy produced in the (muscle) tissue is
needed to overcome the drag from the surrounding fluid. In other words,
the driving force for the traveling waves seen in the nematode has its origins
in the contractions of the muscle.

We model the nematode as a hollow cylindrical shell with outer radius $r_{o}$
and inner radius
$r_{i}$ \cite{park2007} such that the principal moment of inertia (second
moment of the area of cross-section) $I$ along the entire length of the
nematode is given by $I = \frac{\pi}{4}\left[(r_{m} + \frac{t}{2})^{4}
- (r_{m} - \frac{t}{2})^{4}\right]$ where $r_{m}$ is the mean radius and
$t$ is the cuticle thickness.

Consistent with experimental observations, we assume that the nematode moves along the $x$-axis and that the deflections of the centerline of the nematode from the $x$-axis are small. This allows us to take $s=x$ and write $\kappa(x,t) = \partial \phi / \partial x = \partial^{2}y/\partial x^{2}$ where $y(x,t)$ is the deflection of the centerline of the nematode and $\phi(x,t)$ is the angle made
by the tangent $\hat{t}$ to the $x$-axis. The equations of motion can then be written as
\begin{eqnarray}
\frac{\partial F_{x}}{\partial x} = C_{t}V_{x}, &\qquad&
\frac{\partial F_{y}}{\partial x}  = C_{n}\frac{\partial y}{\partial t},
\label{eq:yforbal_app} \\
\frac{\partial M}{\partial x} + F_{y} & =& 0, \label{eq:mombal_app}
\end{eqnarray}
where $\vec{F} = F_{x}\vec{e}_{x} + F_{y}\vec{e}_{y}$. $V_{x}$ is the velocity
of the nematode along the $x$-axis and corresponds to the average forward speed
$U$. Combining a linearized formulation of Eqs.~(\ref{force_app}) and
(\ref{moment_app}) along with the viscoelastic model of Eq.~(\ref{eq:constlaw})
\emph{offers a direct route towards (i) a closed-form analytical solution for
the curvature $\kappa(x, t)$ and (ii) an estimate of tissue properties} ($E$ and
$\eta$). Note that due to the assumption of small
deflections \cite{grayhancock1955}, the $x$-component of the force balance
becomes decoupled from the rest of the equations. As a result, we will not be
able to predict $V_{x}$ even if $y(x,t)$ is determined.
But, we can solve equations (\ref{eq:constlaw}), (\ref{eq:yforbal_app}), and
(\ref{eq:mombal_app}) for appropriate boundary conditions on $F_{y}$ and $M$ to see
if we get solutions that look like traveling waves whose amplitude is not a
constant, but in fact, is decreasing from the head to the tail of the
nematode (Fig.~\ref{fig_2}a). To do so, we observe that the nematode's body oscillates at
a single frequency irrespective of position $x$ along its centerline
(Fig.~\ref{fig_2}b). We therefore assume $y(x,t)$ and $M(x,t)$ to
have a form that involves a single frequency $\omega$, so that
$y(x,t) = f(x)\cos\omega t + g(x)\sin\omega t$, and
\begin{eqnarray}
 M(x,t) &=& K_{b}(\omega)\frac{\partial^{2}f}{\partial x^{2}}
 \cos(\omega t + \psi(\omega)) \nonumber \\
 & & +K_{b}(\omega)\frac{\partial^{2}g}{\partial x^{2}}
 \sin(\omega t + \psi(\omega)), \label{eq:mconst}
\end{eqnarray}
where $f(x)$ and $g(x)$ are as yet unknown functions, $\psi(\omega)$ is
frequency dependent phase angle, and $K_{b}(\omega)$ is a frequency dependent
bending modulus of the homogeneous viscoelastic material making up the
nematode. In particular,
\begin{equation}
 \tan\psi = \frac{\eta \omega}{E}, \qquad
 K_{b} = I\sqrt{E^{2} + \omega^{2}\eta^{2}},
 \label{eq:youngvisco}
\end{equation}
where the parameter $K_b$, $E$, and $\eta$ correspond to the \emph{effective}
tissue properties of the nematode. To be consistent with the observation of a
single frequency $\omega$ and non-zero amplitudes of the curvature
$\kappa$ at the
head and tail, we apply boundary conditions
\begin{equation}
 F_{y}(0,t) = 0, \qquad M(0,t) = 0, \qquad
 F_{y}(L,t) = 0, \qquad M(L,t) = 0,
\end{equation}
We can make further progress by
differentiating the balance of moments once with respect to $x$ and
substituting the $y$-component of the balance of forces into it to get
\begin{equation}
 \frac{\partial^{2}M}{\partial x^{2}}
 + C_{n}\frac{\partial y}{\partial t} = 0
 \label{eq:ode_y_app}
\end{equation}
Substituting for $M(x,t)$ from eqn.(\ref{eq:mconst}) yields a biharmonic
equation for the functions $f$ and $g$ which can be solved to give
the following solution for $y(x,t)$
\begin{eqnarray}
 y(x,t) & =& A_{01}\exp(\beta x\cos(\frac{\pi}{8}+\frac{\psi}{4}))
 \cos(\beta x\sin(\frac{\pi}{8} + \frac{\psi}{4}) - \omega t - \phi_{01})
 \nonumber \\
 & +& A_{23}\exp(-\beta x\cos(\frac{\pi}{8}+\frac{\psi}{4}))
 \cos(-\beta x\sin(\frac{\pi}{8} + \frac{\psi}{4}) - \omega t - \phi_{23})
 \nonumber \\
 & +& A_{45}\exp(\beta x\cos(\frac{\psi}{4}-\frac{3\pi}{8}))
 \cos(\beta x\sin(\frac{\psi}{4} - \frac{3\pi}{8}) - \omega t - \phi_{45})
 \nonumber \\
 & +& A_{67}\exp(-\beta x\cos(\frac{\psi}{4}-\frac{3\pi}{8}))
 \cos(-\beta x\sin(\frac{\psi}{4}-\frac{3\pi}{8}) - \omega t - \phi_{67}),
 \label{eq:yfinal}
\end{eqnarray}
where $\beta = \left(C_{n} \omega/K_{b}\right)^{1/4}$, and $A_{01}$, $A_{23}$,
$A_{45}$, $A_{67}$, $\phi_{01}$, $\phi_{23}$, $\phi_{45}$ and $\phi_{67}$ are
eight constants to be determined from the eight equations resulting from the
$\sin\omega t$ and $\cos\omega t$ coefficients of the boundary conditions. Note
that these are
four waves of the type $y(x,t) = b(x)\cos(2\pi(x + V_{w}t)/\lambda)$ which
is the form originally {\it assumed} by Gray and Hancock based on
experiment \cite{grayhancock1955}. In contrast, we have obtained such
waves as a solution to the equations of motion. We plot the amplitude and phase of these
waves for a particular choice of parameters in Fig.~\ref{fig_6}. But, the exact solution above is cumbersome to use. We need simple expressions that can be easily fit to some
observable in the experiment to obtain $\beta$ and $\psi$, or equivalently,
$E$ and $\eta$. One such parameter is the amplitude of the traveling waves as a function of
position $x$ along the nematode. We develop a strategy to obtain this
amplitude in the following.

Based on the exact solutions to the equations we approximate the
displacement $y(x,t)$ as follows:
\begin{eqnarray}
 y(x,t) & \approx & A_{1}\exp(-\beta x\cos(\frac{\psi}{4} - \frac{3\pi}{8}))
 \cos(-\beta x\sin(\frac{\psi}{4} - \frac{3\pi}{8}) - \omega t) \nonumber \\
 & +& A_{2}\exp(\beta x\cos(\frac{\psi}{4} + \frac{\pi}{8}))
 \cos(\beta x\sin(\frac{\psi}{4} + \frac{\pi}{8}) - \omega t + \phi_{2})
 \nonumber \\
 & +& A_{2}\exp(\beta x\cos(\frac{\psi}{4} - \frac{3\pi}{8}))
 \cos(\beta x\sin(\frac{\psi}{4} - \frac{3\pi}{8}) - \omega t + \phi_{2}
 -\frac{3\pi}{4}) \nonumber \\
 & +& A_{1}\exp(-\beta x\cos(\frac{\psi}{4} + \frac{\pi}{8}))
 \cos(-\beta x\sin(\frac{\psi}{4} + \frac{\pi}{8}) - \omega t - \frac{\pi}{2}).
\end{eqnarray}
For experimental values of $\kappa_{a}(L,t)/\kappa_{a}(0,t) = 0.33$ we find
$A_{1}/A_{2} \approx 443$ so that $A_{2}$ is
much smaller in comparison to $A_{1}$. Note that
\begin{eqnarray}
 y(0,t) & =& A_{1}\cos\omega t + A_{2}\cos(\omega t - \phi_{2})
 + A_{2}\cos(\omega t - \phi_{2} + \frac{3\pi}{4})
 + A_{1}\cos(\omega t +\frac{\pi}{2}) \nonumber \\
 & =& \sqrt{2}A_{1}\cos(\omega t + \frac{\pi}{4})
 +2A_{2}\cos(\frac{3\pi}{8}) \cos(\omega t - \phi_{2} + \frac{3\pi}{8}).
 \label{eq:headdisp}
\end{eqnarray}
The curvature can be calculated as under:
\begin{eqnarray}
 \kappa(x,t) = \frac{\partial^{2}y}{\partial x^{2}} & \approx &
 \beta^{2}A_{1}\exp(-\beta x\cos(\frac{\psi}{4} - \frac{3\pi}{8}))
 \cos(-\beta x\sin(\frac{\psi}{4} - \frac{3\pi}{8}) - \omega t - \frac{3\pi}{4}
 + \frac{\psi}{2}) \nonumber \\
 & +& \beta^{2}A_{2}\exp(\beta x\cos(\frac{\psi}{4} + \frac{\pi}{8}))
 \cos(\beta x\sin(\frac{\psi}{4} + \frac{\pi}{8}) - \omega t + \phi_{2}
 + \frac{\pi}{4} + \frac{\psi}{2})
 \nonumber \\
 & +& \beta^{2}A_{2}\exp(\beta x\cos(\frac{\psi}{4} - \frac{3\pi}{8}))
 \cos(\beta x\sin(\frac{\psi}{4} - \frac{3\pi}{8}) - \omega t + \phi_{2}
 -\frac{3\pi}{4} -\frac{3\pi}{4} + \frac{\psi}{2}) \nonumber \\
 & +& \beta^{2}A_{1}\exp(-\beta x\cos(\frac{\psi}{4} + \frac{\pi}{8}))
 \cos(-\beta x\sin(\frac{\psi}{4} + \frac{\pi}{8}) - \omega t - \frac{\pi}{2}
 + \frac{\pi}{4} + \frac{\psi}{2}).
\end{eqnarray}
Note again that
\begin{eqnarray}
 \kappa(0,t) & =&\beta^{2}A_{1}\cos(\omega t + \frac{3\pi}{4} - \frac{\psi}{2})
 + \beta^{2}A_{2}\cos(\omega t - \phi_{2} - \frac{\pi}{4} - \frac{\psi}{2})
 \nonumber \\
 &  &+ \beta^{2}A_{2}\cos(\omega t - \phi_{2} + \frac{3\pi}{4} + \frac{3\pi}{4}
   -\frac{\psi}{2})
 + \beta^{2}A_{1}\cos(\omega t + \frac{\pi}{4} - \frac{\psi}{2}) \nonumber \\
 & =& \sqrt{2}\beta^{2}A_{1}\cos(\omega t + \frac{\pi}{2} - \frac{\psi}{2})
 + 2A_{2}\cos(\frac{7\pi}{8})\cos(\omega t - \phi_{2} - \frac{\psi}{2}
 + \frac{5\pi}{8}). \label{eq:headcurv}
\end{eqnarray}
It is possible to determine $\beta$ and $\psi$ from (\ref{eq:headdisp})
and (\ref{eq:headcurv}) alone. If we recognize that $A_{1} >> A_{2}$ at the
head then we see that the ratio of the amplitude of the curvature to the
amplitude of the displacement at the head is simply $\beta^{2}$ and the phase
difference between them is $(\frac{\pi}{4} - \frac{\psi}{2})$ (the phase
difference between $\frac{\partial y}{\partial x}$ and
$\frac{\partial^{2}y}{\partial x^{2}}$ at the head is
$\frac{19\pi}{16} - \frac{\psi}{4}$). We find by
comparing with the exact solution that an estimate of $\beta$ using this method
is accurate to within 1\% and that of $\psi$ is accurate to within
2 or 3 degrees.

The curvature is an oscillatory function and we can determine its amplitude
$\mathcal{A}(x)$ simply by isolating the coefficients of $\cos\omega t$ and
$\sin\omega t$ and then squaring and adding them. We get the following expression as a result of this exercise:
\begin{eqnarray}
 \mathcal{A}^{2}(x) & =& \beta^{4}A_{1}^{2}\exp(-2\beta x\cos(\frac{\psi}{4}
 - \frac{3\pi}{8}))
 + \beta^{4}A_{2}^{2}\exp(2\beta x\cos(\frac{\psi}{4} + \frac{\pi}{8}))
 \nonumber \\
 &  &+ \beta^{4}A_{2}^{2}\exp(2\beta x\cos(\frac{\psi}{4} - \frac{3\pi}{8}))
 + \beta^{4}A_{1}^{2}\exp(-2\beta x\cos(\frac{\psi}{4} + \frac{\pi}{8}))
 \nonumber \\
 &  &-2\beta^{4}A_{1}A_{2}\exp(-\sqrt{2}\beta x\sin(\frac{\psi}{4}
 - \frac{\pi}{8})) \cos(-\sqrt{2}\beta x\sin(\frac{\psi}{4}
 - \frac{\pi}{8}) - \phi_{2}) \nonumber \\
 &  &+ 2\beta^{4}A_{1}A_{2}
 \cos(-2\beta x\sin(\frac{\psi}{4} - \frac{3\pi}{8}) -\phi_{2} +\frac{3\pi}{4})
 \nonumber \\
 &  &- 2\beta^{4}A_{2}^{2}\exp(\sqrt{2}\beta x\cos(\frac{\psi}{4}
 - \frac{\pi}{8}))\cos(\sqrt{2}\beta x\cos(\frac{\psi}{4} - \frac{\pi}{8})
 + \frac{3\pi}{4}) \nonumber \\
 &  &+ 2\beta^{4}A_{1}A_{2}
 \cos(2\beta x\sin(\frac{\psi}{4} + \frac{\pi}{8}) + \frac{\pi}{2} + \phi_{2})
 \nonumber \\
 &  &+ 2\beta^{4}A_{1}A_{2}\exp(\sqrt{2}\beta x\sin(\frac{\psi}{4}
 - \frac{\pi}{8}))\cos(\sqrt{2}\beta x\sin(\frac{\psi}{4}
 - \frac{\pi}{8}) + \phi_{2}-\frac{5\pi}{4})
 \nonumber \\
 &  &+ 2\beta^{4}A_{1}^{2}\exp(-\sqrt{2}\beta x\cos(\frac{\psi}{4}
 - \frac{\pi}{8}))\cos(\sqrt{2}\beta x\cos(\frac{\psi}{4} - \frac{\pi}{8})
 -\frac{\pi}{2}).
\end{eqnarray}
We can fit the experimental data of curvature as a function of $x$ using the
expression above. There are five fit parameters -- $A_{1}$, $A_{2}$, $\beta$,
$\psi$ and $\phi_{2}$. The value of $\phi_{2}$ mostly affects the curvature
profile near the tail. Values of $\phi_{2} \approx \pi/3$ seem to give
good fits for the curvature data of the nematodes. $A_{1}$ can be determined
from the amplitude of the displacement
at the head. This leaves three fit parameters -- $A_{2}$, $\beta$ and $\psi$.
We can use this fit to check if the parameters $\beta$ and $\psi$ obtained from
analyzing the motion of the head alone are reasonable or not.

\subsection*{Methods Summary}
\textbf{\emph{C. elegans} strain}. All strains were maintained using standard culture methods and fed with the \emph{E. coli} strain OP50. The following muscular dystrophic (MD) strains were used: \emph{fer-1(hc24ts)}, \emph{dys-1(cx18)I} and \emph{hlh-1(cc561)II;dys-1(cx18)I} double mutant. Note that \emph{hlh-1} is a myoD mutant that qualitatively reveals the motility defects of \emph{dys-1} mutants. Analysis were performed on hypochlorite synchronized young adult animals. \emph{fer-1} mutants were hatched at the restrictive temperature of $25^o$C and grown until they reach the young adult stage. \emph{dys-1(cx18) I}; \emph{hlh-1(cc561) II} mutants were grown at the permissive temperature of $16^o$C. Wild-type nematodes grown at the appropriate temperature were used as controls. Strains were obtained from the Caenorhabditis elegans Genetic Stock Center.

\subsection*{Acknowledgements}
The authors would like to thank J. Yasha Kresh, Y. Goldman, P. Janmey, and T. Shinbrot for helpful discussions. We also thank R. Sznitman for help with vision algorithms and P. Rockett for manufacturing acrylic channels. Some nematode strains used in this work were provided by the Caenorhabditis Genetics Center, which is funded by the NIH National Center for Research Resources (NCRR).

\newpage
\section*{Figure Legends}
\begin{figure*}[p]
\begin{center}
\includegraphics[width=0.9\textwidth]{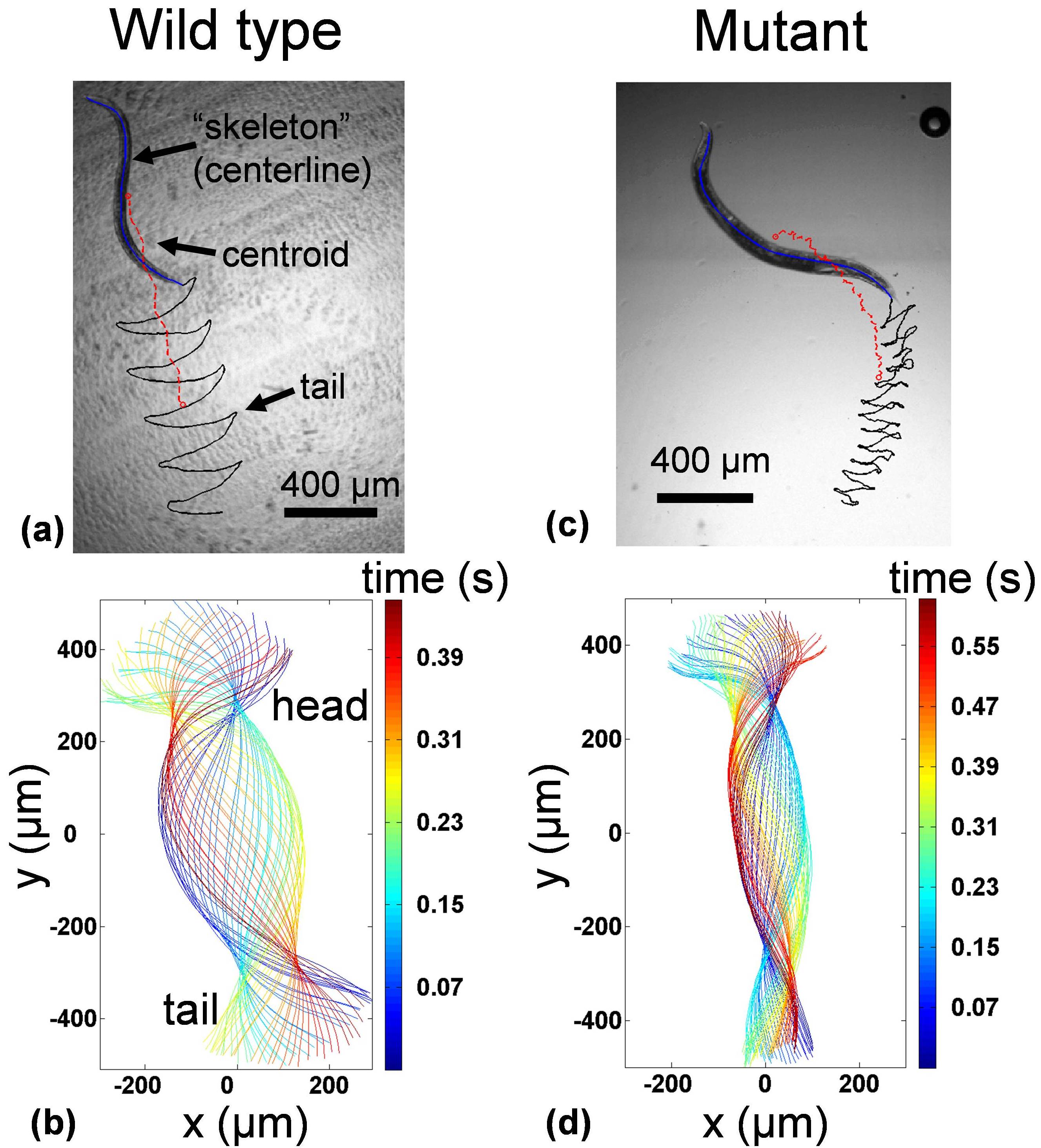}
\caption{}
\label{fig_1}
\end{center}
\end{figure*}

~\\\textbf{Figure~\ref{fig_1}:}\\
\textbf{Motility of wild-type \emph{C. elegans} and \emph{dys-1;hlh-1} muscular dystrophic (MD) mutant swimming at low Reynolds number}. \textbf{(a)} and \textbf{(c)}: Visualization of \emph{C. elegans} motion illustrating instantaneous body centerline or skeleton. Also shown are the nematode's (i) centroid and (ii) tail-tip trajectories over multiple body bending cycles. \textbf{(b)} and \textbf{(d)}: Color-coded temporal evolution of \emph{C. elegans} skeletons over one beating cycle. Results reveal a well-defined envelope of elongated body shapes with a wavelength corresponding approximately to the nematode's body length.

\begin{figure*}[p]
\begin{center}
\includegraphics[width=0.9\textwidth]{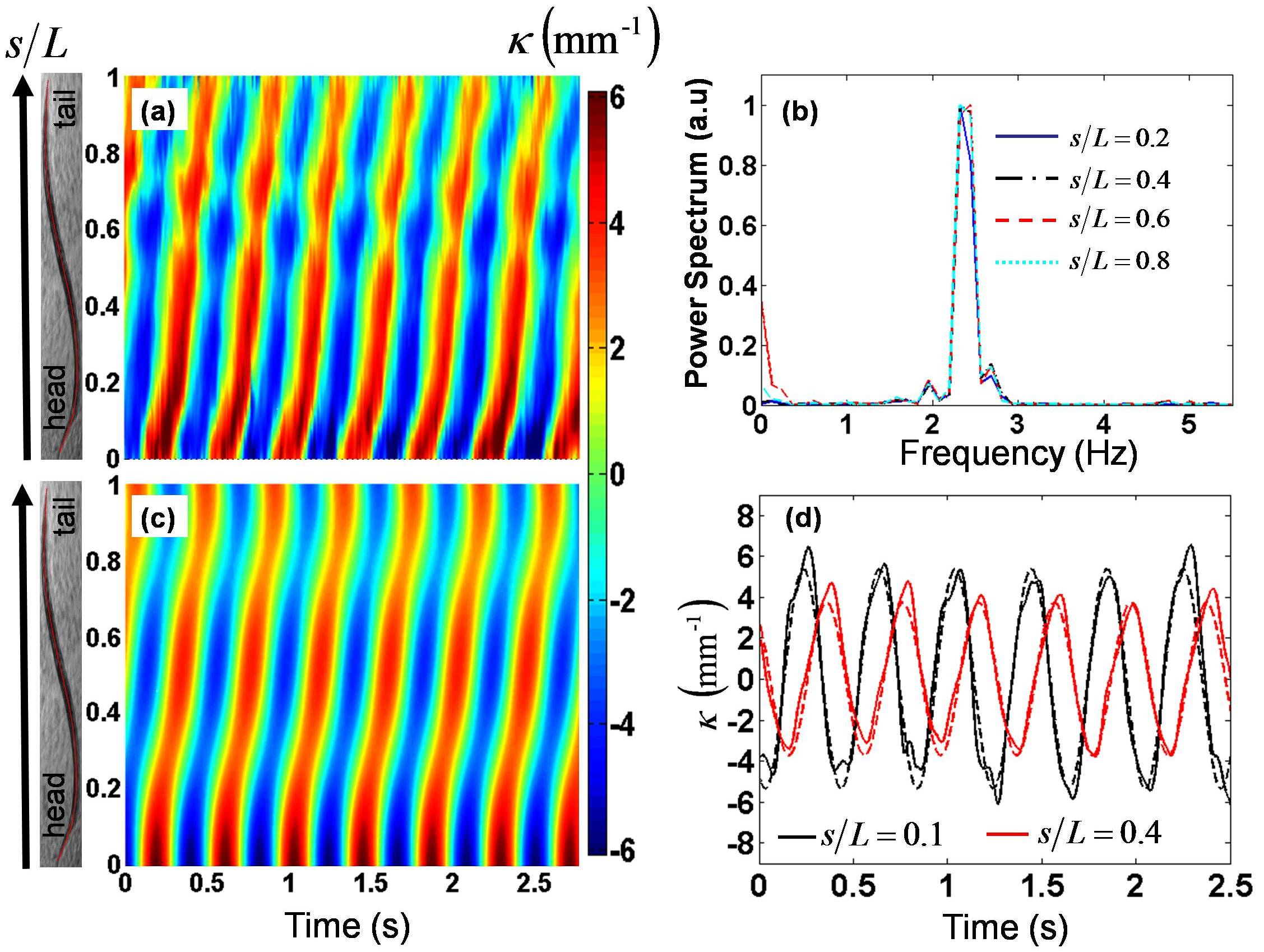}
\caption{}
\label{fig_2}
\end{center}
\end{figure*}

~\\\textbf{Figure~\ref{fig_2}:}\\
\textbf{Spatio-temporal kinematics of \emph{C. elegans} forward swimming gait}. (\textbf{a}) Representative contour plot of the \emph{experimentally measured} curvature ($\kappa$) along the nematode's body centerline for approximately $6$ bending cycles. Red and blue colors represent positive and negative $\kappa$  values, respectively. The $y$-axis corresponds to the dimensionless position $s/L$ along the \emph{C. elegans}' body length where $s=0$ is the head and $s=L$ is the tail. (\textbf{b}) Nematode's body bending frequency obtained from Fast Fourier Transform of $\kappa$ at different $s/L$. The peak is seen at a single frequency ($\sim 2.4$~Hz) irrespective of the location $s/L$. \textbf{(c)} Contour plot of curvature $\kappa$ values \emph{obtained from the model}. The model captures the longitudinal bending wave with decaying magnitude, which travels from head to tail. (\textbf{d}) Comparison between experimental and theoretical curves of $\kappa$ at  $s/L=0.1$ and $s/L=0.4$; dashed lines correspond to model predictions (root mean square error is $\sim 10\%$ of peak-to-peak amplitude).

\begin{figure*}[p]
\begin{center}
\includegraphics[width=0.7\textwidth]{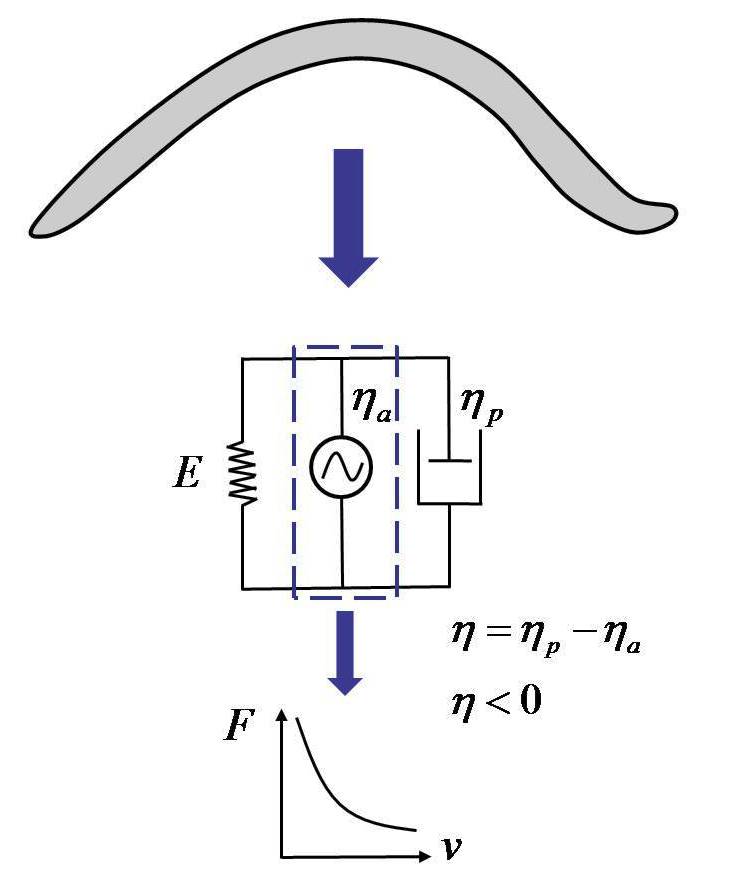}
\caption{}
\label{fig_3}
\end{center}
\end{figure*}

~\\\textbf{Figure~\ref{fig_3}:}\\
\textbf{Schematic of the analytical model for the total internal moment $M$}. Muscle tissue is described by a visco-elastic model containing both passive and active elements. The passive moment ($M_p$) is described by the Voigt model consisting of a passive elastic element (spring) of stiffness $E$ (i.e. Young's modulus) and a passive viscous element (dashpot) of tissue viscosity $\eta_p$. The active moment ($M_a$) is described by an active muscular element of viscosity $\eta_a$ and illustrates a negative slope on a force-velocity plot. Since there is a net generation of energy in the muscle to overcome drag from the surrounding fluid, we expect $\eta = \eta_p - \eta_a < 0$.

\begin{figure*}[p]
\begin{center}
\includegraphics[width=0.9\textwidth]{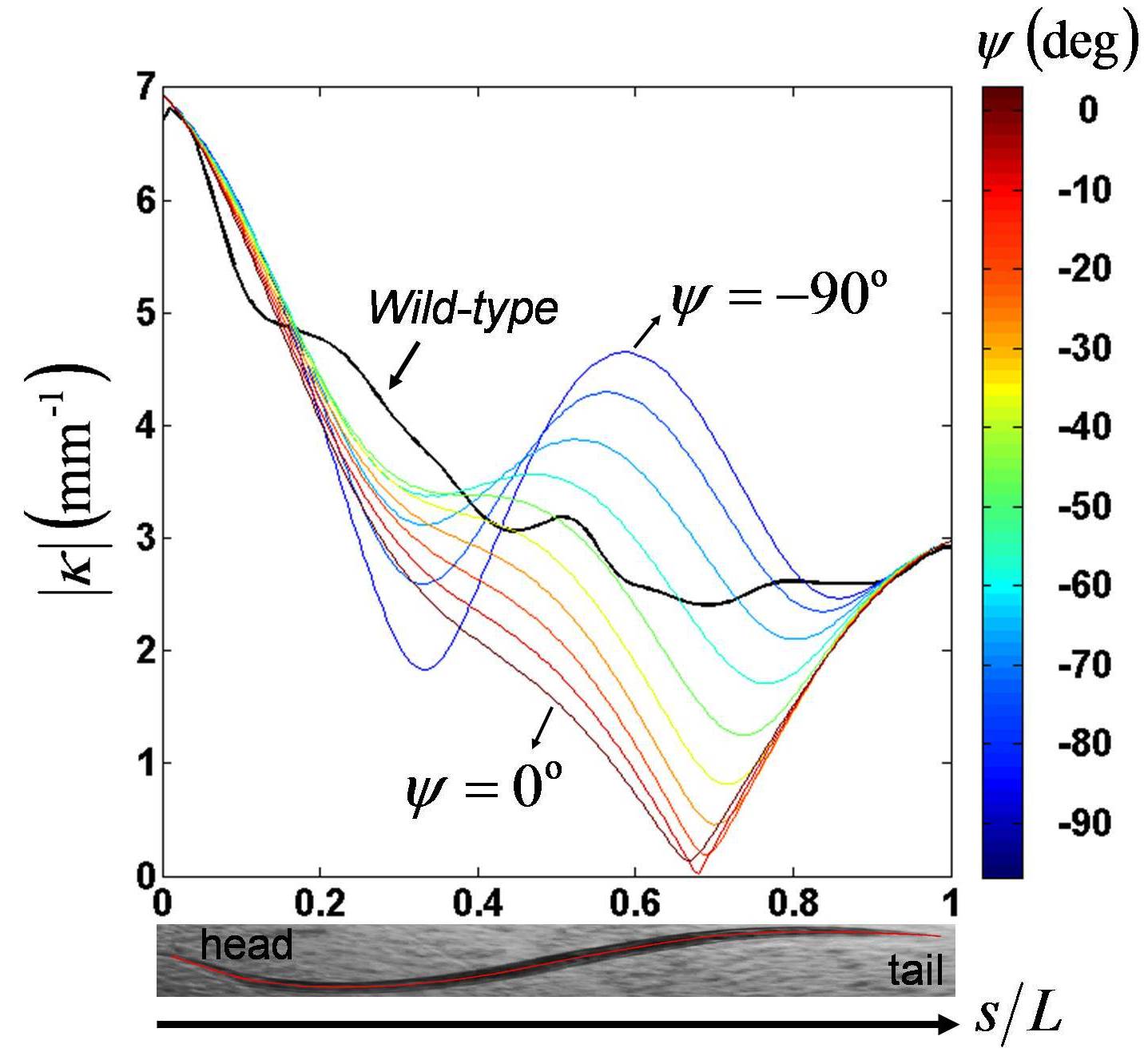}
\caption{}
\label{fig_4}
\end{center}
\end{figure*}

~\\\textbf{Figure~\ref{fig_4}:}\\
\textbf{Typical \emph{C. elegans} viscoelastic material properties}. Typical experimental profile of the curvature amplitude $|\kappa|$ decay as a function of body position $s/L$. Color-coded theoretical profiles of the curvature amplitude decay $|\kappa|$ at fixed value of the bending modulus $K_b$. Curves vary from $\psi = 0^o$ (red) to $\psi = -90^o$ (blue), which corresponds to $\eta=0$ and $E=0$, respectively.

\begin{figure*} [p]
\begin{center}
\includegraphics[width=1.0\textwidth]{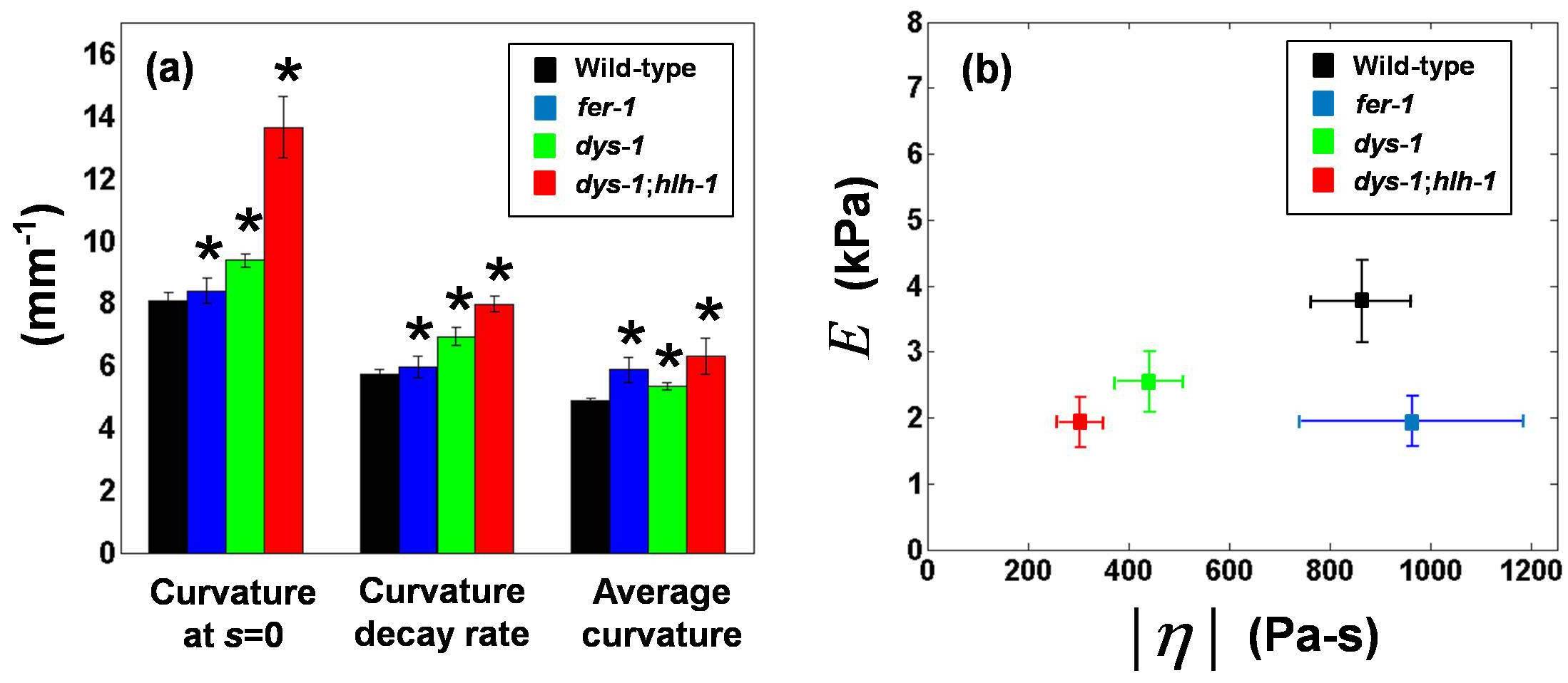}
\caption{}
\label{fig_5}
\end{center}
\end{figure*}

~\\\textbf{Figure~\ref{fig_5}:}\\
\textbf{Kinematics and material properties of wild type and three muscle mutants of \emph{C. elegans}}. (\textbf{a}) Measured kinematic data and (\textbf{b}) estimated Young's modulus $E$ and absolute values of tissue viscosity $|\eta|$ for wild-type, \emph{fer-1(hc24)}, \emph{dys-1(cx18)}, and \emph{dys-1(cx18)}; \emph{hlh-1(cc561)} adult nematodes ($n = 7$-$25$ nematodes for each genotype.  * - $p <0.01$).

\begin{figure*} [p]
\begin{center}
\includegraphics[width=1.0\textwidth]{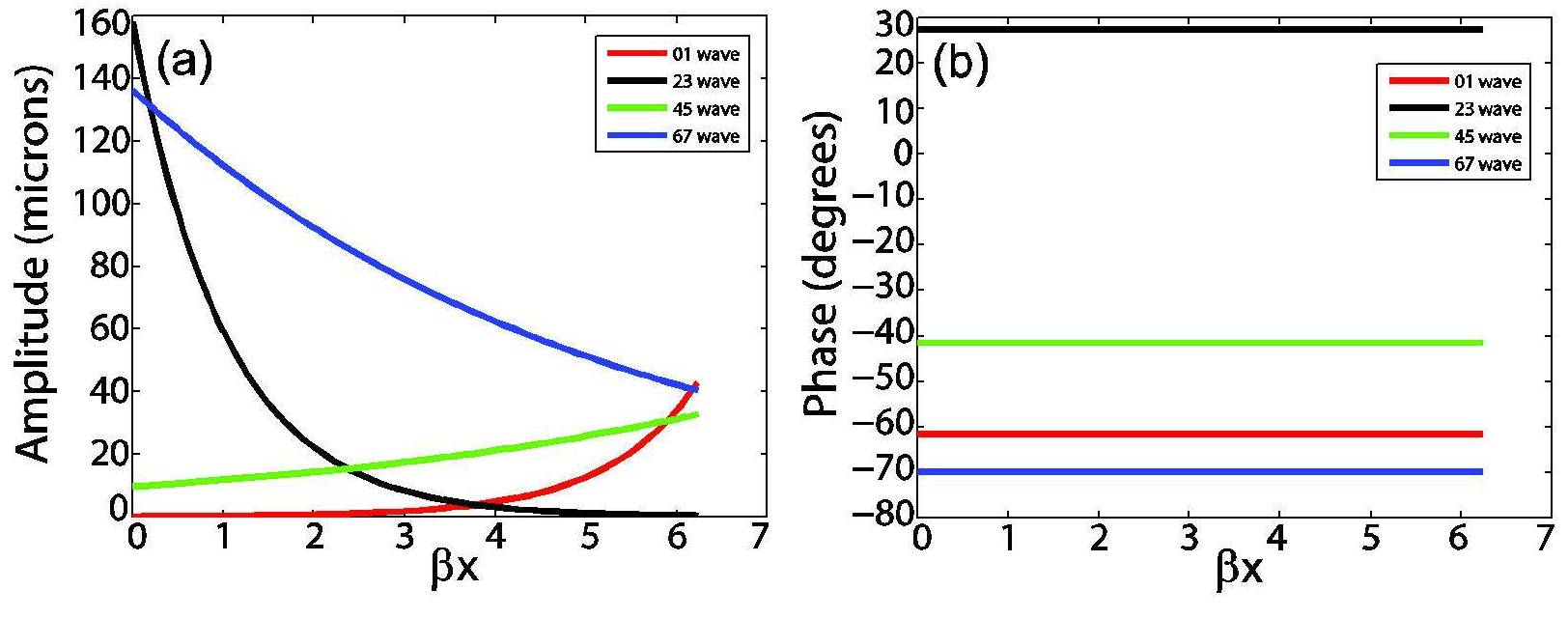}
\caption{}
\label{fig_6}
\end{center}
\end{figure*}

~\\\textbf{Figure~\ref{fig_6}:}\\
\textbf{Amplitude and phase of traveling waves obtained from enforcing
force and moment boundary conditions at $x=0,L$}. (\textbf{a}) Amplitudes
of the $A_{23}$ and $A_{67}$ waves decrease from $x=0$ to $x=L$
while the amplitudes of the $A_{01}$ and $A_{45}$ waves increase from $x=0$ to $x=L$. The amplitudes of the former two waves are
larger than the latter two. (\textbf{b}) Phases $\phi_{01}$, $\phi_{23}$,
$\phi_{45}$ and $\phi_{67}$ are all constant from $x=0$ to $x=L$. The phase difference between the $A_{67}$ wave and $A_{23}$ wave is approximately $90^o$. The parameters used to obtain these plots are:
$\omega = 4\pi$ radians/sec, $L = 1.0$mm, $K_{b} = 5.0\times 10^{-16}$Nm$^{2}$, $C_{n} = 0.06$Ns/m$^{2}$, $\psi =-45^o$, $EIQ_{0} = 4.35\times 10^{-12}$Nm,
$Q_{1}L =-1.054Q_{0}$ and $B = 198.4^{o}$.

\newpage

\newpage

\newpage

\newpage

\pagebreak
\newpage

\pagebreak
\newpage
\newpage

\end{document}